\documentclass[showpacs,prd]{revtex4}
\usepackage{epsfig}
\usepackage{graphicx}
\usepackage{dcolumn}
\usepackage{amsmath}
\usepackage{latexsym}

\begin{document} 
%%%%%%%%%%%%%%%%%%%%%%%%%%%%%%%%%%%%%%%%%%%%%%%%%%%%%%%%%%%%%%%%%%%%%%%%%%%%
%%%%%%%%%%%%%%%%%%%%%%%%%%%%%%%%%%%%%%%%%%%%%%%%%%%%%%%%%%%%%%%%%%%%%%%%%%%%

\begin{flushright}
NITheP-08-10\\
\end{flushright}

\title{Formulation, Interpretation and Application of non-Commutative Quantum Mechanics}  
\date{\today}
\author{F G Scholtz$^{a,b}$\footnote{E-mail: fgs@sun.ac.za}, L Gouba $^b$\footnote{E-mail:
gouba@sun.ac.za}, A Hafver$^{a}$, C M Rohwer$^{a}$}
\affiliation{$^a$Institute of Theoretical Physics, University of Stellenbosch, Stellenbosch 7600, South Africa\\
$^b$National Institute for Theoretical Physics (NITheP), Stellenbosch Institute of Advanced Study, Stellenbosch 7600, South Africa}

\begin{abstract}
In analogy with conventional quantum mechanics, non-commutative quantum mechanics is formulated as a quantum system on the Hilbert space of Hilbert-Schmidt operators acting on non-commutative configuration space. It is argued that the standard quantum mechanical interpretation based on Positive Operator Valued Measures, provides a sufficient framework for the consistent interpretation of this quantum system. The implications of this formalism for rotational and time reversal symmetry are discussed. The formalism is applied to the free particle and harmonic oscillator in two dimensions and the physical signatures of non commutativity are identified.  
\end{abstract}

\maketitle

\section{Introduction}
\label{intro}
There seems to be growing consensus that our notion of space-time has to be drastically revised in a consistent formulation of quantum mechanics and gravity \cite{dop},\cite{seib}.  One possible generalization, suggested by string theory \cite{wit}, that has attracted much interest recently is that of non-commutative space-time \cite{doug}.  Despite a number of investigations into the possible physical consequences of non-commutativity in quantum mechanics and quantum mechanical many-body systems \cite{li},\cite{mendes}, \cite{bem}, \cite{khan}, quantum electrodynamics \cite{chai},\cite{chair}, \cite{lia}, the standard model \cite{ohl} and cosmology  \cite{gar}, \cite{alex}), our understanding of the physical implications of non-commutativity is still in its infancy and even controversial \cite{fiore}. Part of the difficulty in developing a thorough understanding of the physical implications of non- commutativity, is the lack of a systematic formulation and interpretational framework of non-commutative quantum mechanics.  Indeed, even a simple system such as a particle confined to a spherical well presents a challenge as it is difficult to give meaning to the notion of a well in non-commutative space.  This problem was solved only recently \cite{fgs} by using the notion of projection operators on non-commutative configuration space to define piecewise constant potentials.  Here we develop the notions of \cite{fgs} into a fully fledged formulation and interpretational framework for non-commutative quantum mechanics.  

For simplicity we confine the discussion here to two dimensional systems.  The extension to three dimensions is more involved due to the necessity to identify the correct representation of the rotational symmetry on both the non-commutative configuration space and the quantum Hilbert space. These issues will be discussed in a forthcoming publication \cite{fgs1}.    

In section \ref{Formalism}, we develop the formalism and interpretational framework of non-commutative quantum systems.  This includes the derivation of a continuity equation and the conservation of probability.  In section \ref{symmetries} we discuss the implications of this formalism for rotational and time reversal symmetry.  In section \ref{applications} we apply this formalism to the free particle problem and the harmonic oscillator and identify the physical consequences of non commutativity.  Finally, we close in section \ref{conclusions} with a discussion, conclusion and future perspective.  

\section{Formalism}
\label{Formalism}

We develop the formalism of non-commutative quantum mechanics in complete analogy with commutative quantum mechanics.  To fully exploit this analogy, it is useful to briefly review the wave mechanical formulation (position representation) of commutative quantum mechanics.  A particle moving in $d$-dimensions is described in wave mechanics by a configuration space $R^d$ and a Hilbert space $L^2$ of square integrable wave functions $\psi(x)$ over $R^d$.  The inner product on $L^2$ is 
$\left(\phi,\psi\right)=\int d^dx\phi^*(x)\psi(x)$.  We denote the elements of this Hilbert space by $\psi(x)\equiv|\psi\rangle$ and the elements of its dual (linear functionals) by $\langle\psi|$, which maps elements of $L^2$ onto complex numbers by $\langle\phi|\psi\rangle=\left(\phi,\psi\right)$.  The basic axioms of commutative quantum mechanics are

\begin{enumerate}

\item A physical state of the system is represented by a density operator on $L^2$, which is a positive definite Hermitian operator with trace one. The density operator of a pure state is a projection operator $\rho=|\psi\rangle\langle\psi|$. 

\item Physical observables correspond to Hermitian operators $A$ on $L^2$.

\item The outcomes of a measurement of an observable $A$ are the eigenvalues $a$ of $A$.

\item Given that the system is in a physical state described by the density operator $\rho$, the probability of finding $a$ in a measurement of the observable $A$ is $P(a)=\langle a|\rho|a\rangle$ with $|a\rangle$ the corresponding normalised eigenstate of $A$.

\item After the measurement the system is in the state $\rho=|a\rangle\langle a|$ and an immediate repeated measurement of $A$ yields $a$ with probability one, consistent with the hermiticity of $A$ and $(4)$.  This is often referred to as von Neuman's projection postulate.    

\end{enumerate}

A key element in the actual construction of the quantum system above, as well as for the identification of the observables, is to find a unitary representation of the abstract Heisenberg algebra
\begin{eqnarray}
\label{heis}
\left[x_i,p_j\right]&=&i\hbar\delta_{i,j},\\\nonumber
\left[x_i,x_j\right]&=&0,\\
\left[p_i,p_j\right]&=&0.\nonumber
\end{eqnarray}
in terms of operators $\hat{x}_i$ and $\hat{p}_i$ acting on the space of square integrable functions.  This is just the well known Schr\"odinger representation
\begin{equation}
\label{sch}
\hat x_i\psi(x)=x_i\psi(x),\quad \hat p_i\psi(x)=-i\hbar\frac{\partial\psi(x)}{\partial x_i}
\end{equation}
which acts irreducibly and, from the Stone- von Neumann theorem, is known to be unique up to unitary transformations.     
We take this mathematical structure and interpretational framework as our cue for developing the mathematical structure and interpretational framework of the non-commutative system.            

We start by considering non-commutative configuration space.  Restricting to two dimensions, the coordinates of non-commutative configuration space satisfy the commutation relation 
\begin{equation}
[\hat{x}_i, \hat{x}_j] = i\theta\epsilon_{i,j},
\end{equation} 
with  $\theta$ a real positive parameter and $\epsilon_{i,j}$ the completely anti-symmetric tensor with $\epsilon_{1,2}=1$. It is convenient to define the creation and annihilation operators
\begin{eqnarray}\nonumber
b &=& \frac{1}{\sqrt{2\theta}} (\hat{x}_1+i\hat{x}_2),\\
b^\dagger &=&\frac{1}{\sqrt{2\theta}} (\hat{x}_1-i\hat{x}_2),
\end{eqnarray}
that satisfy the Fock algebra $[ b ,b^\dagger ] = 1$. The non-commutative configuration space is then isomorphic to boson Fock space
\begin{eqnarray}
\mathcal{H}_c = \textrm{span}\{ |n\rangle\equiv \frac{1}{\sqrt{n!}}(b^\dagger)^n |0\rangle\}_{n=0}^{n=\infty},
\end{eqnarray}
where the span is taken over the field of complex numbers.

The next step is to introduce the equivalent of the Hilbert space of square integrable function in which the physical states of the system are to be represented.  This mathematical structure is actually well known and a natural generalization of the space of square integrable functions.  We consider the set of Hilbert-Schmidt operators acting on non-commutative configuration space
\begin{equation}
\label{qhil}      
\mathcal{H}_q = \left\{ \psi(\hat{x}_1,\hat{x}_2): \psi(\hat{x}_1,\hat{x}_2)\in \mathcal{B}\left(\mathcal{H}_c\right),\;{\rm tr_c}(\psi(\hat{x}_1,\hat{x}_2)^\dagger\psi (\hat{x}_1,\hat{x}_2) < \infty \right\}.
\end{equation}
Here ${\rm tr_c}$ denotes the trace over non-commutative configuration space and $\mathcal{B}\left(\mathcal{H}_c\right)$ the set of bounded operators on $\mathcal{H}_c$.  This space has a natural inner product and norm 
\begin{equation}\label{inner}
\left(\phi(\hat{x}_1,\hat{x}_2),\psi(\hat{x}_1,\hat{x}_2)\right) = {\rm tr_c}(\phi(\hat{x}_1,\hat{x}_2)^\dagger\psi(\hat{x}_1,\hat{x}_2)).
\end{equation}
and forms a Hilbert space \cite{hol}. This space is the analog of the space of square integrable wave functions of commutative quantum mechanics and to distinguish it from the non-commutative configuration space $\mathcal{H}_c$, which is also a Hilbert space, we shall refer to it as quantum Hilbert space and use the footnotes $c$ and $q$ to make this distinction. We also use the notation $|\cdot\rangle$ for elements of non-commutative configuration space, while elements of the quantum Hilbert space are denoted by $\psi(\hat{x},\hat{y})\equiv |\psi)$ and the elements of its dual (linear functionals) by $(\psi|$, which maps elements of $\mathcal{H}_q$ onto complex numbers by $\left(\phi|\psi\right)=\left(\phi,\psi\right)={\rm tr_c}\left(\phi(\hat{x}_1,\hat{x}_2)^\dagger\psi(\hat{x}_1,\hat{x}_2)\right)$.  We also need to be careful when denoting Hermitian conjugation to distinguish between these two spaces. We reserve the notation $\dagger$ to denote Hermitian conjugation on non-commutative configuration space and the notation $\ddagger$ for Hermitian conjugation on quantum Hilbert space.  

The abstract Heisenberg algebra is now replaced by the non-commutative Heisenberg algebra. In two dimensions this reads
\begin{eqnarray}
\label{heisnc}
\left[{x}_i,{p}_j\right] &=& i\hbar\delta_{i,j},\nonumber\\
\left[{x}_i,{x}_j\right] &=& i\theta\epsilon_{i,j},\\
\left[{p}_i,{p}_j\right] &=& 0\nonumber.
\end{eqnarray}
A unitary representation of this algebra in terms of operators $\hat{X}_i$ and $\hat{P}_i$ acting on the quantum Hilbert space (\ref{qhil}) with inner product (\ref{inner}), which is the analog of the Schr\"{o}dinger representation of the Heisenberg algebra, is easily found to be 
\begin{eqnarray}
\label{schnc}
\hat{X}_i\psi(\hat{x}_1,\hat{x}_2) &=& \hat{x}_i\psi(\hat{x}_1,\hat{x}_2),\nonumber\\
\hat{P}_i\psi(\hat{x}_1,\hat{x}_2) &=& \frac{\hbar}{\theta}\epsilon_{i,j}[\hat{x}_j,\psi(\hat{x}_1,\hat{x}_2)],
\end{eqnarray}
i.e., the position acts by left multiplication and the momentum adjointly.  We use capital letters to distinguish operators acting on quantum Hilbert space from those acting on non-commutative configuration space. It is also useful to introduce the following quantum operators 
\begin{eqnarray}
\label{qop}
B&=&\frac{1}{\sqrt{2\theta}}\left(\hat{X}_1+i\hat{X}_2\right),\nonumber\\
B^\ddagger&=&\frac{1}{\sqrt{2\theta}}\left(\hat{X}_1-i\hat{X}_2\right),\nonumber\\
\hat{P}&=&\hat{P}_1 + i\hat{P}_2,\nonumber\\
\hat{P}^\ddagger &=& \hat{P}_1 -i\hat{P}_2.
\end{eqnarray}
We note that $\hat{P}^2=\hat{P}^2_1+\hat{P}^2_2 = P^\ddagger P = PP^\ddagger$.  These operators act as follow
\begin{eqnarray}
\label{aqo}
B\psi(\hat{x}_1,\hat{x}_2) &=& b\psi(\hat{x}_1,\hat{x}_2),\nonumber\\
B^\ddagger\psi(\hat{x}_1,\hat{x}_2) &=& b^\dagger\psi(\hat{x}_1,\hat{x}_2),\nonumber\\
P\psi(\hat{x}_1,\hat{x}_2)&=& -i\hbar \sqrt{\frac{2}{\theta}}[b,\psi(\hat{x}_1,\hat{x}_2)],\nonumber\\
P^\ddagger\psi(\hat{x}_1,\hat{x}_2) &=& i\hbar\sqrt{\frac{2}{\theta}}[ b^{\dagger},\psi(\hat{x}_1,\hat{x}_2)].
\end{eqnarray}

We now take the axioms of commutative quantum mechanics to apply with the simple replacement of $L^2$ by $\mathcal{H}_q$.  Although this provides a consistent interpretational framework, the measurement of position needs more careful consideration. The problem with a measurement of position is not that the axioms above do not apply to the Hermitian operators $\hat{X}_i$, rather the problem is that these operators do not commute and thus a precise measurement of one of these observables leads to total uncertainty in the other. Yet, we would like to preserve the notion of position in the sense of a particle being localized around a certain point.  The best we can do in the non-commutative situation is to construct a minimal uncertainty state in non-commutative configuration space and use that to give meaning to the notion of position. The issue of position measurement in non-commutative space was also considered in \cite{kempf}, but the approach described below is different and utilizes the standard notion of Operator Valued Measures.    

The minimal uncertainty states on non-commutative configuration space, which is isomorphic to boson Fock space, are well known to be the normalized coherent states \cite{klaud}
\begin{equation}
\label{cs} 
|z\rangle = e^{-z\bar{z}/2}e^{z b^{\dagger}} |0\rangle,
\end{equation}
where $z=\frac{1}{\sqrt{2\theta}}\left(x_1+ix_2\right)$ is a dimensionless complex number.  These states provide an overcomplete basis on the non-commutative configuration space.  Corresponding to these states we can construct a state (operator) in quantum Hilbert space as follows
\begin{equation}
|z )=|z\rangle\langle z|.
\end{equation}
Writing the trace in terms of coherent states and using $|\langle z|w\rangle|^2=e^{-|z-w|^2}$ one easily verifies that this is indeed a Hilbert-Schmidt operator.  These states also have the property
\begin{equation}
B|z)=z|z),
\end{equation}
which leads to the natural interpretation of $\left(x_1,x_2\right)$ as the dimensionful position coordinates.  Another property of these states is that they provide an overcomplete set on quantum Hilbert space and a resolution of the identity in the form
\begin{equation}
{1}_q=\int \frac{dz d\bar{z}}{\pi} |z)e^{\stackrel{\leftarrow}{\partial_{\bar{z}}}\stackrel{\rightarrow}{\partial_z}}(z|=\int \frac{dx_1 dx_2}{2\pi\theta} |z)e^{\stackrel{\leftarrow}{\partial_{\bar{z}}}\stackrel{\rightarrow}{\partial_z}}(z|,
\end{equation}
with $\partial_{\bar z}\equiv \frac{\partial}{\partial \bar{z}}$ and $\partial_z\equiv \frac{\partial}{\partial z}$.  To prove this identity consider a Hilbert-Schmidt operator $\psi$ acting on non-commutative configuration space, which is an element of quantum Hilbert space denoted by $\psi=|\psi)$.  It follows that $\left(z|\psi\right)=\langle z|\psi|z \rangle$. Next we make use of the well known result that any operator $\psi$ on boson Fock space (non-commutative configuration space) can be written as \cite{klaud} 
\begin{eqnarray}
\psi=|\psi)&=&\int \frac{dz d\bar{z}}{\pi}\;|z\rangle\langle z|\left[ e^{-\partial_{\bar{z}}\partial_z}\langle z|\psi|z\rangle\right],\nonumber\\
&=&\int \frac{dz d\bar{z}}{\pi}\;|z) e^{-\partial_{\bar{z}}\partial_z}\left(z|\psi\right),\nonumber\\
&=&\int \frac{dz d\bar{z}}{\pi}\;|z) e^{\stackrel{\leftarrow}{\partial_{\bar{z}}}\stackrel{\rightarrow}{\partial_z}}\left(z|\psi\right).
\end{eqnarray}  
In the last step the exponent was expanded and a term by term partial integration performed, keeping in mind that boundary terms vanish due to the Gaussian factor in the coherent state.  The result now follows immediately.

With these results in place we can set up an interpretational framework for the measurement of position.  The first thing we note is that the states $|z)$ are actually eigenstates of a non-Hermitian operator and thus non orthogonal.  This is the essential difficulty with position measurements in non-commutative quantum mechanics.  However, a formalism to deal with this situation, referred to as Positive Operator Valued Measures has been developed in the context of open quantum systems \cite{ber}.  Introduce the dimensionful operators 
\begin{equation}
\label{povm}
\pi_z=\frac{1}{2\pi\theta}|z) e^{\stackrel{\leftarrow}{\partial_{\bar{z}}}\stackrel{\rightarrow}{\partial_z}}(z|\;,\quad \int dx_1 dx_2\;\pi_z=1_q\,.
\end{equation}
These operators are Hermitian on quantum Hilbert space, positive and they integrate to the identity.  They therefore provide an Operator Valued Measure in the sense of \cite{ber} and we can give a consistent probability interpretation by assigning the probability of finding the particle at position $\left(x_1,x_2\right)$, given that the system is described by the pure state density matrix $\rho=|\psi ) (\psi|$, to be 
\begin{eqnarray}
P(x_1,x_2)={\rm tr_q}\left(\pi_z\rho\right)=\left(\psi|\pi_z|\psi\right)\geq 0,\\\nonumber
\int dx_1dx_2 P(x_1,x_2)=\int dx_1 dx_2\left(\psi|\pi_z|\psi\right)=\left(\psi|\psi\right)=1,
\end{eqnarray}
where we assumed the states $|\psi )$ to be normalized. The only difference with the conventional probability interpretation is that the operators $\pi_z$ are not orthogonal as would be the case when working with Hermitian operators. This implies that we can no longer make a statement about the final state after a measurements as an immediate repetition of the measurement may yield a different result.  This requires us, as is done in Positive Operator Valued Measures \cite{ber}, to relax the von Neumann projection axiom (axiom $(4)$) by not postulating a unique state of the system after a measurement of position. Rather one proceeds as follows:  Since the operators $\pi_z$ are positive, we can write them as $\pi_z=A_z^\ddagger A_z$.  Clearly the operators $A_z$, referred to as the detection operators, are not uniquely defined and any choice of the form $A_z=U_z\pi_z^{1/2}$, with $U_z$ unitary is acceptable. The von Neumann projection postulate then gets replaced by \cite{ber}

\begin{itemize}

\item After the measurement the system is in the state $|\phi\rangle=\frac{1}{\sqrt{\langle\psi|A^\ddagger_z A_z|\psi\rangle}}A_z|\psi\rangle$ if it was initially in the pure state $|\psi\rangle$ and $\rho_z=\frac{A_z\rho A_z^\ddagger}{{\rm tr}\left(A^\ddagger_z A_z\rho\right)}$ if it was initially in the mixed state $\rho$.

\end{itemize}
The final state may therefore involve an arbitrary unitary transformation over which we have no control and which reflects the inherent uncertainty of a non-commutative world.    

A final point to check is the conservation of probability.  We must show that the norm of the state is preserved under time evolution as in the commutative case. To be specific, we consider a Hamiltonian acting on quantum Hilbert space 
\begin{equation}
H=\frac{P^2}{2m}+V\left(\hat{X}_1,\hat{X}_2\right).
\end{equation}
We assume that the potential $V\left(\hat{x}_1,\hat{x}_2\right)$, viewed as an operator acing on non-commutative configuration space, is Hermitian, i.e., $V^\dagger\left(\hat{x}_1,\hat{x}_2\right)=V\left(\hat{x}_1,\hat{x}_2\right)$ (the equivalent of requiring the potential to be real in commutative quantum mechanics). The time dependent Schr\"{o}dinger equation reads
\begin{equation}
i\hbar\frac{\partial\psi\left(\hat{x}_1,\hat{x}_2,t\right)}{\partial t}=H\psi\left(\hat{x}_1,\hat{x}_2,t\right).    
\end{equation}
Recalling that the operators $\psi\left(\hat{x}_1,\hat{x}_2,t\right)$ act on non-commutative configuration space, we can multiply this equation from the left with $\psi^\dagger\left(\hat{x}_1,\hat{x}_2,t\right)$ and its Hermitian conjugate, taken w.r.t. the inner product on configuration space, by $\psi\left(\hat{x}_1,\hat{x}_2,t\right)$.  Subtracting the two equations we obtain the analog of the continuity equation
\begin{equation}
\frac{\partial}{\partial t}
\rho=[\hat{x}_2,j_{1}]+[\hat{x}_1,j_{2}],\label{eq: continuity equation}
\end{equation}
with 
\begin{eqnarray}
\rho&=&\psi^{\dagger}\left(\hat{x}_1,\hat{x}_2,t\right)\psi\left(\hat{x}_1,\hat{x}_2,t\right),\nonumber\\ j_{1}&=&\frac{\hbar}{2mi\theta^{2}}(\psi^{\dagger}\left(\hat{x}_1,\hat{x}_2,t\right)[\hat{x}_2,\psi\left(\hat{x}_1,\hat{x}_2,t\right)]-[\hat{x}_2,\psi^{\dagger}\left(\hat{x}_1,\hat{x}_2,t\right)]\psi\left(\hat{x}_1,\hat{x}_2,t\right)),\nonumber\\
j_{2}&=&\frac{\hbar}{2mi\theta^{2}}(\psi^{\dagger}\left(\hat{x}_1,\hat{x}_2,t\right)[\hat{x}_1,\psi\left(\hat{x}_1,\hat{x}_2,t\right)]-[\hat{x}_1,\psi^{\dagger}\left(\hat{x}_1,\hat{x}_2,t\right)]\psi\left(\hat{x}_1,\hat{x}_2,t\right)).
\end{eqnarray}

Tracing this equation over configuration space, the right hand side vanishes as it is a commutator and we assumed Hilbert-Schmidt operators (equivalent of square integrability for which the boundary terms vanish) and we conclude $\frac{\partial}{\partial t} {\rm tr_c}\left(\psi^\dagger\psi\right)=\frac{\partial}{\partial t}\left(\psi|\psi\right)=0$, 
which is the statement of probability conservation. 

The formalism and interpretational framework is now in place and in the next sections we explore the consequences of this formalism further.

\section{Symmetries}
\label{symmetries}

We start with the rotational symmetry in the plane.  We are looking for an operator that generates rotations in the non-commutative configuration space.  This is easily found to be the operator
\begin{equation}
\ell_z=\frac{-i}{2\theta}\left(\hat{x}_1^2+\hat{x}_2^2\right).
\end{equation}
It satisfies the commutation relations
\begin{eqnarray}
\left[\ell_z,\hat{x}_1\right]&=&-\hat{x}_2,\\\nonumber
\left[\ell_z,\hat{x}_2\right]&=&\hat{x}_1,
\end{eqnarray}
i.e., it generates rotations around the z-axis.  Setting $U=e^{\phi\ell_z}$ finite rotations are obtained by
\begin{eqnarray}
\hat{x}_1^\prime&=&U^\dagger\hat{x}_1 U=\hat{x}_1\cos\phi+\hat{x}_2\sin\phi,\nonumber\\
\hat{x}_2^\prime&=&U^\dagger\hat{x}_2 U=\hat{x}_2\cos\phi-\hat{x}_1\sin\phi.
\end{eqnarray}

In analogy with commutative quantum mechanics, we can now derive the transformation a rotation on the non-commutative configuration space induces on the quantum Hilbert space:
\begin{equation}
\psi\left(\hat{x}^\prime_1,\hat{x}^\prime_2\right)=\psi\left(U^\dagger \hat{x}_1 U,U^\dagger \hat{x}_2 U,\right)=U^\dagger\psi\left(\hat{x}_1,\hat{x}_2\right)U.
\end{equation} 
For infinitesimal rotation $\delta\phi$ this reads
\begin{eqnarray}
\psi\left(\hat{x}^\prime_1,\hat{x}^\prime_2\right)&=&\psi\left(\hat{x}_1,\hat{x}_2\right)+\delta\phi\left[\psi\left(\hat{x}_1,\hat{x}_2\right),\ell_z\right]\nonumber\\
&=&\psi\left(\hat{x}_1,\hat{x}_2\right)+\delta\phi\left(\frac{-i}{\hbar}\right)\left(\hat{X}_1\hat{P}_2-\hat{X}_2\hat{P}_1+\frac{\theta}{2\hbar}\hat{P}^2\right)\psi\left(\hat{x}_1,\hat{x}_2\right)\\\nonumber
&=&e^{\left(\frac{-i\delta\phi}{\hbar}\right) L_z}\psi\left(\hat{x}_1,\hat{x}_2\right),
\end{eqnarray}
where we have identified the quantum angular momentum operator as
\begin{equation}
L_z=\left(\hat{X}_1\hat{P}_2-\hat{X}_2\hat{P}_1+\frac{\theta}{2\hbar}\hat{P}^2\right).
\end{equation}

The next issue we discuss is time reversal symmetry.  In commutative quantum mechanics the antiunitary time reversal operator $\Theta$ corresponds to complex conjugation.  In the non-commutative case one easily infers from the time dependent Sch\"{o}dinger equation that $\Theta$ must be Hermitian conjugation on non-commutative configuration space, i.e.,
\begin{equation}
\Theta\psi\left(\hat{x}_1,\hat{x}_2\right)=\psi^\dagger\left(\hat{x}_1,\hat{x}_2\right).
\end{equation}
From this one infers the relations
\begin{eqnarray}
\label{trev}
\Theta \hat{X}_i\Theta^{-1}&=&\hat{X}_i+
\frac{\theta}{\hbar}\epsilon_{i,j}\hat{P}_j,\nonumber\\
\Theta\hat{P}_i\Theta^{-1}&=&-\hat{P}_i,\nonumber\\
\Theta L_z\Theta^{-1}&=&-L_z.
\end{eqnarray}
The first relation immediately implies that for non-constant potentials $\Theta H\Theta^{-1}\ne H$ and thus time reversal symmetry breaking.  A detailed example of this is worked out when the formalism is applied to the harmonic oscillator.

\section{Applications}
\label{applications}
\subsection{Free particle}

Let us consider the Schr\"{o}dinger equation for a free particle 
\begin{equation}\label{sche}
i\hbar\frac{\partial}{\partial t} |\psi,t) = \hat{H }|\psi,t),
\end{equation}
where
\begin{equation}
\hat{H} = \frac{\hat{P}^2}{2m}.
\end{equation}
As usual the time dependence can be factored out
\begin{eqnarray}
|\psi,t) = e^{-iEt/\hbar} |\psi),
\end{eqnarray}
where the time independent wave function $|\psi)$ satisfies the time independent Schr\"{o}dinger equation
\begin{equation}
\frac{\hat{P}^2}{2m} |\psi) = E |\psi).
\end{equation}

The solution is a non-commutative plane wave
\begin{equation}
\label{pwa}
|\psi) = e^{\theta |k|^2}e^{i\theta k b}e^{i\theta \bar{k}b^\dagger},
\end{equation}
where $k$ is a complex number and the energy dependent part of the divergent normalization $\left(\psi|\psi\right)={\rm tr_c}e^{-\theta|k|^2}$ has been included.  The energy is given by
\begin{equation}
E = \frac{\hbar^2\theta|k|^2}{m}.
\end{equation}
According to our interpretation of \ref{Formalism} the probability to find the particle at $z=\frac{1}{\sqrt{2\theta}}\left(x_1+ix_2\right)$ is 
\begin{eqnarray}
P(x_1,x_2) \propto (\psi|z)e^{\stackrel{\leftarrow}{\partial}_{\bar{z}}
\stackrel{\rightarrow}{\partial}_{z}}(z|\psi).
\end{eqnarray}
Using the Hadamard formula for two operators $A$ and $B$
\begin{eqnarray}
e^B A e^{-B} = A + [B,A] +\frac{1}{2!}[B,[B,A]] +\frac{1}{3!}[B,[B,[B,A]]]
\ldots
\end{eqnarray}
the exponentials can be re-ordered to yield
\begin{eqnarray}
(z|\psi)=e^{-\theta^2|k|^2}e^{i\theta k z + i\theta\bar{k}\bar{z}}.
\end{eqnarray}
We now find that the probability $P(x_1,x_2)$ is independent of $(x_1,x_2)$ as expected for a free particle.

\subsection{Harmonic oscillator}

The Hamiltonian of the non-commutative harmonic oscillator is
\begin{equation}
\hat{H} = \frac{1}{2m}\hat{P}_1^2 + \frac{1}{2m}\hat{P}_2^2 + \frac{1}{2}m\omega^2\hat{X}_1^2 + \frac{1}{2}m\omega^2\hat{X}_2^2,
\end{equation}
where $m$ and $\omega$ are, respectively, the mass and frequency. 

It is convenient to rewrite the Hamiltonian as follows
\begin{equation}\label{ham}
\hat{H} = \frac{1}{2m}\left({\hat{P}_1}^2 + {\hat{P}_2}^2 + {\hat{X'}_1}^2 + {\hat{X'}_2}^2\right)
\end{equation}
where $\hat{X}_1' = m\omega \hat{X}_1$ and $\hat{X}_2' = m\omega\hat{X}_2$ satisfy the algebra
\begin{eqnarray}
\left[\hat{X}_1',\hat{X}_2'\right]  &=& im^2\omega^2\theta\,,\nonumber\\
\left[\hat{X}_1',\hat{P_1}\right] &=& im\omega\hbar\,,\nonumber\\
\left[\hat{X}_2',\hat{P_2}\right]  &=& im\omega\hbar\,,\nonumber\\
\left[\hat{P}_1,\hat{P}_2\right] &=& 0.
\end{eqnarray}
The natural procedure to follow from here is to introduce creation and annihilation operators that diagonalize the Hamiltonian as in the commutative case.  Set
\begin{equation}
2m\hat{H} =Z^{T}Z,
\end{equation}
where the transposed vector $Z^T = \left(\hat{X}_1',\hat{X}_2',\hat{P}_1,\hat{P}_2\right)$. Creation and annihilation operators $V^T=\hat{A}_1,\hat{A}_1^\ddagger, \hat{A}_2, \hat{A}_2^\ddagger$ are introduced as linear combinations
\begin{eqnarray}
V=SZ
\end{eqnarray}

These operators should satisfy the Fock algebra 
\begin{eqnarray}
\label{fock}
\left[ \hat{A}_i,\hat{A}_j^\ddagger \right] &=& \delta_{i,j}.
\end{eqnarray}

Define ${V^\ddagger}^T=\left(\hat{A}_1^\ddagger,\hat{A}_1, \hat{A}_2^\ddagger, \hat{A}_2\right)=V^T\Lambda$ with
\begin{eqnarray}
\Lambda = \left(
\begin{array}{llll}
0 & 1 & 0 & 0\\
1 & 0 & 0 & 0\\
0 & 0 & 0 & 1 \\
0 & 0 & 1 & 0
\end{array}\right).
\end{eqnarray}

We have $V^\ddagger=S^*Z$, which yields from (\ref{fock})

\begin{equation}
\label{scond}
SgS^\dagger=D 
\end{equation}
where 
\begin{eqnarray}
D = \left(
\begin{array}{llll}
1 & 0 & 0 & 0\\
0 & -1 & 0 & 0\\
0 & 0 & 1 & 0 \\
0 & 0 & 0 & -1
\end{array}\right)
\end{eqnarray}
and
\begin{eqnarray}
g_{i,j}\equiv\left[Z_i,Z_j\right] = \left(
\begin{array}{llll}
0 & im^2\omega^2\theta & i\hbar m\omega & 0\\
-im^2\omega^2\theta & 0 & 0 & i\hbar m\omega\\
-i\hbar m\omega & 0 & 0 & 0 \\
0 & -i\hbar m\omega & 0 & 0
\end{array}\right).
\end{eqnarray}

It follows that $S^{-1}=DgS^\dagger$ and the Hamiltonian becomes
\begin{eqnarray}
\label{hamsec}
2mH&=&Z^TZ\nonumber\\
&=&\left(V^\ddagger\right)^T\left(S^\dagger\right)^{-1}S^{-1}V\nonumber\\
&=&\left(V^\ddagger\right)^T S g^2S^\dagger V.
\end{eqnarray}

Since $g$ is Hermitian and $g^*=-g$, its eigenvalues are real and come in two pairs $\left(\lambda_1,-\lambda_1\right)$ and $\left(\lambda_2,-\lambda_2\right)$, $\lambda_1,\lambda_2\ge 0$, and the corresponding eigenvectors $\left(u_1^+,u_1^-\right)$ and $\left(u_2^+,u_2^-\right)$ are orthogonal.  Setting the columns of $S^\dagger$ to be the normalized eigenvectors of $g$ divided by the square root of the absolute value of the corresponding eigenvalue, i.e.
\begin{equation}
S^\dagger=\left(\frac{u_1^+}{\sqrt\lambda_1},\frac{u_1^-}{\sqrt\lambda_1},\frac{u_2^+}{\sqrt\lambda_2},\frac{u_2^-}{\sqrt\lambda_2}\right),
\end{equation}
we solve (\ref{scond}) and the Hamiltonian becomes
\begin{equation}
\label{fhe}
\hat{H} = \frac{\lambda_1}{2m}(2\hat{A}_1^\ddagger \hat{A}_1 +1) + \frac{\lambda_2}{2m}(2\hat{A}_2^\ddagger\hat{A}_2 +1).
\end{equation}
The explicit form of $\lambda_1$, $\lambda_2$ are
\begin{eqnarray}
\label{lam}
\lambda_1 &=& 
\frac{1}{2}\left(m^2\omega^2\theta + m\omega\sqrt{4\hbar^2 + m^2\omega^2\theta^2}\right)\,,\nonumber\\
\lambda_2 &=& 
\frac{1}{2}\left(-m^2\omega^2\theta + m\omega\sqrt{4\hbar^2 + m^2\omega^2\theta^2}\right),
\end{eqnarray}
while the creation and annihilation operators are given by
\begin{eqnarray}
\label{crann}
\hat{A}_1 &=& \frac{1}{\sqrt{K_1}}
\left(-\frac{\lambda_1}{\hbar}\hat{X}_1 - i\frac{\lambda_1}{\hbar}\hat{X}_2
-i\hat{P}_1 +\hat{P}_2\right),\nonumber\\
\hat{A}_1^\ddagger &=&  \frac{1}{\sqrt{K_1}}
\left(-\frac{\lambda_1}{\hbar}\hat{X}_1 + i\frac{\lambda_1}{\hbar}\hat{X}_2
+i\hat{P}_1 +\hat{P}_2\right),\nonumber\\
\hat{A}_2 &=& \frac{1}{\sqrt{K_2}}
\left(\frac{\lambda_2}{\hbar}\hat{X}_1 - i\frac{\lambda_2}{\hbar}\hat{X}_2
+i\hat{P}_1 +\hat{P}_2\right),\\
\hat{A}_2^\ddagger &=&\frac{1}{\sqrt{K_2}}
\left(\frac{\lambda_2}{\hbar}\hat{X}_1 +i\frac{\lambda_2}{\hbar}\hat{X}_2
-i\hat{P}_1 +\hat{P}_2\right),
\end{eqnarray}
with
\begin{eqnarray}
\label{norm}
K_1&=&\lambda_1\left(\frac{2\lambda_1\theta}{\hbar^2}+4\right),\nonumber\\
K_2&=&\lambda_2\left(-\frac{2\lambda_2\theta}{\hbar^2}+4\right).
\end{eqnarray}
In the commutative limit ($\theta=0$) this reduces to the usual result as $\lambda_1=\lambda_2=\hbar m\omega$.
  
The next step is to determine the ground state wave function $\psi_0$, which must satisfy the following conditions
\begin{eqnarray}\label{wav}
\hat A_1\psi_0 = 0 ,\\\label{wavv}
\hat A_2\psi_0 = 0.
\end{eqnarray}
We make the ansatz 
\begin{equation}
\label{grd}
\psi_0 = e^{\alpha b^\dagger b},
\end{equation}
where $\alpha $ is to be determined.  From (\ref{crann}) the condition 
\begin{equation}
\hat A_1\psi_0 = 0,
\end{equation} 
reads 
\begin{eqnarray}
\left(\frac{\lambda_1}{\hbar}(\hat{X}_1+i\hat{X}_2)
-i(\hat{P}_1+i \hat{P}_2)\right)\psi_0 = 0
\end{eqnarray} 
or, equivalently,
\begin{equation}
\label{alp}
\frac{\lambda_1}{\hbar}\sqrt{2\theta}
b\psi_0 -\hbar\sqrt{\frac{2}{\theta}}[b,\psi_0]  = 0.
\end{equation}

Using
\begin{equation}\label{int}
[b,\psi_0] = (1-e^{-\alpha}) b \psi_0,
\end{equation}
we have
\begin{eqnarray}
\label{aeq}
\left(\frac{\lambda_1}{\hbar}\sqrt{2\theta}
-\hbar\sqrt{\frac{2}{\theta}}(1-e^{-\alpha})\right) b \psi_0 = 0,
\end{eqnarray}
and thus  
\begin{equation}
\label{al1}
e^{-\alpha} = 1+ \frac{\theta}{\hbar^2}\lambda_1.
\end{equation}
The same procedure applied to 
\begin{equation}
\hat A_2\psi_0 = 0
\end{equation} 
gives
\begin{equation}
\label{al2}
e^{\alpha} = 1 -\frac{\theta}{\hbar^2}\lambda_2.
\end{equation}
From (\ref{lam}) it is easily verified that (\ref{al1}) and (\ref{al2}) are consistently solved by 
\begin{eqnarray}\nonumber
\alpha = \ln \left(1-\frac{\theta}{\hbar^2}\lambda_2\right)=-\ln \left(1+\frac{\theta}{\hbar^2}\lambda_1\right).
\end{eqnarray}

Using $\hat{x}_1^2+\hat{x}_2^2=\theta\left(2b^\dagger b+1\right)$, the ground state wave function can be written, up to an irrelevant constant, as
\begin{equation}
\psi_0 = e^{\frac{\alpha}{2\theta}(\hat{x}_1^2 + \hat{x}_2^2)}.
\end{equation}
In the commutative limit this yield the standard result
\begin{equation}
\psi_0 = e^{-\frac{m\omega}{2\hbar}(\hat{x}_1^2 + \hat{x}_2^2)}.
\end{equation}

The creation and annihilation operators transform as follow under a rotation
\begin{eqnarray}
\left[L_z,\hat{A}_1\right] &=& \hbar\hat{A}_1,\nonumber\\
\left[L_z, \hat{A}_1^\ddagger\right] &=& -\hbar\hat{A}_1^\ddagger,\nonumber\\
\left[L_z, \hat{A}_2^\ddagger\right] &=& \hbar\hat{A}_2^\ddagger, \nonumber\\
\left[L_z, \hat{A}_2\right] &=& -\hbar \hat{A}_2,
\end{eqnarray}
from which it follows that that $L_z$ also commutes with the Hamiltonian. Clearly $A_1^\ddagger$ creates states with angular momentum $-1$, while $A_2^\ddagger$ creates states with angular momentum $+1$ in units of $\hbar$ 

The angular momentum of the ground state is easy to find 
\begin{eqnarray}
L_z\psi_0  &=& (\hat{X}_1\hat{P}_2 -\hat{X}_2\hat{P}_1 +\frac{\theta}{2\hbar}\hat{P}_1^2 +\frac{\theta}{2\hbar}\hat{P}_2^2)\psi_0\nonumber\\
          &=&-\frac{\hbar}{\theta}\hat{x}_1[\hat{x}_1,\psi_0] -\frac{\hbar}{\theta}\hat{x}_2[\hat{x}_2,\psi_0]
          + \frac{\hbar}{2\theta}[\hat{x}_2,[\hat{x}_2,\psi_0]]
          + \frac{\hbar}{2\theta}[\hat{x}_1,[\hat{x}_2,\psi_0]]\nonumber\\
         &=&-\frac{\hbar}{2\theta}[\hat{x}_1^2+\hat{x}_2^2,\psi_0]=0.
\end{eqnarray}
Therefore, as expected, the ground state carries zero angular momentum.  The full spectrum, including the angular momentum of states, is now obtained by acting with $A_1^\ddagger$ and $A_2^\ddagger$ on the ground state.

Using (\ref{lam}) and the relation $\sqrt{\frac{K_2}{K_1}}=\frac{\lambda_2}{\lambda_1}$ following from it, one verifies that the creation and annihilation operators transform as follow under time reversal:
\begin{eqnarray}
\Theta \hat{A}_1\Theta^{-1} &=&-\hat{A}_2,\nonumber\\
\Theta \hat{A}_1^\ddagger\Theta^{-1} &=& -\hat{A}_2^\ddagger,\nonumber\\
\Theta \hat{A}_2\Theta^{-1} &=&-\hat{A}_1,\nonumber\\
\Theta \hat{A}_2^\ddagger\Theta^{-1} &=& -\hat{A}_1^\ddagger,\nonumber\\
\end{eqnarray}
As $\lambda_1\ne\lambda_2$ for $\theta\ne 0$, it follows from these relations that $\Theta H\Theta^{-1}\ne H$ and thus that time reversal symmetry is broken when non commutativity is present.

The probability to find the particle at position $z=\frac{1}{\sqrt{2\theta}}\left(x_1+ix_2\right)$ is given by
\begin{equation}
P(x_1,x_2)= \left(\psi_0|\pi_z|\psi_0\right) \propto (\psi_0|z)e^{\frac{\stackrel{\leftarrow}{\partial}}{\partial\bar{z}}
\frac{\stackrel{\rightarrow}{\partial}}{\partial z}}(z|\psi_0).
\end{equation}
We have
\begin{eqnarray}
\label{fin}
\left( z|\psi_0\right)=\left( \psi_0|z\right) &=& \langle z |\psi_0| z\rangle\\
&=&e^{-z\bar{z}/2}\langle z|e^{e^\alpha z b^\dagger}|0\rangle\\
&=&e^{-(1 - e^\alpha)|z|^2} = e^{-\frac{\lambda_2\theta}{\hbar^2}|z|^2}.   
\end{eqnarray}
This yields for the probability
\begin{eqnarray}
P(x_1,x_2) &\propto& e^{[(\frac{\lambda_2\theta}{\hbar^2})^2-2\frac{\lambda_2\theta}{\hbar^2} ]\left(x_1^2+x_2^2\right)}.
\end{eqnarray}
In the commutative limit this yields the standard result
\begin{eqnarray}
P(x_1,x_2) \propto e^{\frac{-m\omega}{\hbar}(x_1^2 +x_2^2)}.
\end{eqnarray}
Another interesting limit is the sharply confining potential when $\omega\rightarrow\infty$. In this case one finds
\begin{eqnarray}
P(x_1,x_2) \propto e^{-\frac{1}{2\theta}(x_1^2 +x_2^2)}.
\end{eqnarray}
In contrast to the commutative case where the probability tends to a Dirac delta function in this limit, the non-commutative case yields a Gaussian localized on the scale of $\theta$.  This is to be expected as the non commutativity does not allow localization on a smaller scale.

\section{Conclusion}
\label{conclusions}

We have established a consistent formulation and interpretational framework of non-commutative quantum mechanics.  This framework was demonstrated for the case of a free particle and the harmonic oscillator.  The results found conform to one's expectations of the behaviour of a non-commutative system and in all cases reduce to the standard commutative result in the limit $\theta=0$. Most of the results presented here for the free particle and harmonic oscillator are of course known in the literature \cite{1}-\cite{7}.  However, the aim here was to give a consistent formulation and  interpretational framework for non-commutative quantum mechanics, which includes an unambiguous description for position measurement. Indeed, the result for the position probability distribution in the ground state of the harmonic oscillator is, as far as we can establish, new.  An added advantage that follows from the current formulation is the straightforward extension to much more difficult potentials such as the spherical well \cite{fgs} and the unambiguous formulation of the path integral representation for the propagator \cite{sun}.

\noindent{\bf Acknowledgements.}
This work was supported under a grant of the  National Research Foundation of South Africa. 

%%%%%%%%%%%%%%%%%%%%%%%%%%%%%%%%%%%%%%%%%%%%%%%%%%%%%%%%%%%%%%%%%%%%%%%%%%%

%%%%%%%%%%%%%%%%%%%%%%%%%%%%%%%%%%%%%%%%%%%%%%%%%%%%%%%%%%%%%%%%%%%%%%%%%%%%
%%%%%%%%%%%%%%%%%%%%%%%%%%%%%%%%%%%%%%%%%%%%%%%%%%%%%%%%%%%%%%%%%%%%%%%%%%%%
\end{document}